\newread\epsffilein    
\newif\ifepsffileok    
\newif\ifepsfbbfound   
\newif\ifepsfverbose   
\newdimen\epsfxsize    
\newdimen\epsfysize    
\newdimen\epsftsize    
\newdimen\epsfrsize    
\newdimen\epsftmp      
\newdimen\pspoints     
\pspoints=1bp          
\epsfxsize=0pt         
\epsfysize=0pt         
\def\epsfbox#1{\global\def\epsfllx{72}\global\def\epsflly{72}%
   \global\def\epsfurx{540}\global\def\epsfury{720}%
   \def\lbracket{[}\def\testit{#1}\ifx\testit\lbracket
   \let\next=\epsfgetlitbb\else\let\next=\epsfnormal\fi\next{#1}}%
\def\epsfgetlitbb#1#2 #3 #4 #5]#6{\epsfgrab #2 #3 #4 #5 .\\%
   \epsfsetgraph{#6}}%
\def\epsfnormal#1{\epsfgetbb{#1}\epsfsetgraph{#1}}%
\def\epsfgetbb#1{%
%
%
\openin\epsffilein=#1
\ifeof\epsffilein\errmessage{I couldn't open #1, will ignore it}\else
%
%
   {\epsffileoktrue \chardef\other=12
    \def\do##1{\catcode`##1=\other}\dospecials \catcode`\ =10
    \loop
       \read\epsffilein to \epsffileline
       \ifeof\epsffilein\epsffileokfalse\else
%
%
          \expandafter\epsfaux\epsffileline:. \\%
       \fi
   \ifepsffileok\repeat
   \ifepsfbbfound\else
    \ifepsfverbose\message{No bounding box comment in #1; using defaults}\fi\fi
   }\closein\epsffilein\fi}%
%
%
\def\epsfsetgraph#1{%
   \epsfrsize=\epsfury\pspoints
   \advance\epsfrsize by-\epsflly\pspoints
   \epsftsize=\epsfurx\pspoints
   \advance\epsftsize by-\epsfllx\pspoints
%
%
   \epsfxsize\epsfsize\epsftsize\epsfrsize
   \ifnum\epsfxsize=0 \ifnum\epsfysize=0
      \epsfxsize=\epsftsize \epsfysize=\epsfrsize
%
%
     \else\epsftmp=\epsftsize \divide\epsftmp\epsfrsize
       \epsfxsize=\epsfysize \multiply\epsfxsize\epsftmp
       \multiply\epsftmp\epsfrsize \advance\epsftsize-\epsftmp
       \epsftmp=\epsfysize
       \loop \advance\epsftsize\epsftsize \divide\epsftmp 2
       \ifnum\epsftmp>0
          \ifnum\epsftsize<\epsfrsize\else
             \advance\epsftsize-\epsfrsize \advance\epsfxsize\epsftmp \fi
       \repeat
     \fi
   \else\epsftmp=\epsfrsize \divide\epsftmp\epsftsize
     \epsfysize=\epsfxsize \multiply\epsfysize\epsftmp   
     \multiply\epsftmp\epsftsize \advance\epsfrsize-\epsftmp
     \epsftmp=\epsfxsize
     \loop \advance\epsfrsize\epsfrsize \divide\epsftmp 2
     \ifnum\epsftmp>0
        \ifnum\epsfrsize<\epsftsize\else
           \advance\epsfrsize-\epsftsize \advance\epsfysize\epsftmp \fi
     \repeat     
   \fi
%
%
   \ifepsfverbose\message{#1: width=\the\epsfxsize, height=\the\epsfysize}\fi
   \epsftmp=10\epsfxsize \divide\epsftmp\pspoints
   \vbox to\epsfysize{\vfil\hbox to\epsfxsize{%
      \includegraphics{#1}%
      \hfil}}%
\epsfxsize=0pt\epsfysize=0pt}%

%
%
{\catcode`\%=12 \global\let\epsfpercent=
%
%
\long\def\epsfaux#1#2:#3\\{\ifx#1\epsfpercent
   \def\testit{#2}\ifx\testit\epsfbblit
      \epsfgrab #3 . . . \\%
      \epsffileokfalse
      \global\epsfbbfoundtrue
   \fi\else\ifx#1\par\else\epsffileokfalse\fi\fi}%
%
%
\def\epsfgrab #1 #2 #3 #4 #5\\{%
   \global\def\epsfllx{#1}\ifx\epsfllx\empty
      \epsfgrab #2 #3 #4 #5 .\\\else
   \global\def\epsflly{#2}%
   \global\def\epsfurx{#3}\global\def\epsfury{#4}\fi}%
%
%
\def\epsfsize#1#2{\epsfxsize}
%
%

%
%
%
\def\unredoffs{} \def\redoffs{\voffset=-.31truein\hoffset=-.465truein}
\def\speclscape{}
%
%
%
%
%
%
\newbox\leftpage \newdimen\fullhsize \newdimen\hstitle \newdimen\hsbody
\tolerance=1000\hfuzz=2pt
\catcode`\@=11 
\def\bigans{b }
\def\answ{b }
\ifx\answ\bigans\message{(This will come out unreduced.}
\magnification=1200\unredoffs\baselineskip=16pt plus 2pt minus 1pt
\hsbody=\hsize \hstitle=\hsize 
\else\message{(This will be reduced.} \let\l@r=L
\magnification=1000\baselineskip=16pt plus 2pt minus 1pt \vsize=7truein
\redoffs \hstitle=8truein\hsbody=4.75truein\fullhsize=10truein\hsize=\hsbody
\output={\ifnum\pageno=0 
  \shipout\vbox{\speclscape{\hsize\fullhsize\makeheadline}
    \hbox to \fullhsize{\hfill\pagebody\hfill}}\advancepageno
  \else
  \almostshipout{\leftline{\vbox{\pagebody\makefootline}}}\advancepageno 
  \fi}
\def\almostshipout#1{\if L\l@r \count1=1 \message{[\the\count0.\the\count1]}
      \global\setbox\leftpage=#1 \global\let\l@r=R
 \else \count1=2
  \shipout\vbox{\speclscape{\hsize\fullhsize\makeheadline}
      \hbox to\fullhsize{\box\leftpage\hfil#1}}  \global\let\l@r=L\fi}
\fi
%
\newcount\yearltd\yearltd=\year\advance\yearltd by -1900

\def\Title#1#2{\nopagenumbers\abstractfont\hsize=\hstitle\rightline{#1}%
\vskip 1in\centerline{\titlefont #2}\abstractfont\vskip .5in\pageno=0}
\def\Date#1{\vfill\leftline{#1}\tenpoint\supereject\global\hsize=\hsbody%
\footline={\hss\tenrm\folio\hss}}
%

\def\draftmode{\message{ DRAFTMODE }\def\draftdate{{\rm preliminary draft:
\number\month/\number\day/\number\yearltd\ \ \hourmin}}%
\headline={\hfil\draftdate}\writelabels\baselineskip=20pt plus 2pt minus 2pt
 {\count255=\time\divide\count255 by 60 \xdef\hourmin{\number\count255}
  \multiply\count255 by-60\advance\count255 by\time
  \xdef\hourmin{\hourmin:\ifnum\count255<10 0\fi\the\count255}}}
\def\nolabels{\def\wrlabeL##1{}\def\eqlabeL##1{}\def\reflabeL##1{}}
\def\writelabels{\def\wrlabeL##1{\leavevmode\vadjust{\rlap{\smash%
{\line{{\escapechar=` \hfill\rlap{\sevenrm\hskip.03in\string##1}}}}}}}%
\def\eqlabeL##1{{\escapechar-1\rlap{\sevenrm\hskip.05in\string##1}}}%
\def\reflabeL##1{\noexpand\llap{\noexpand\sevenrm\string\string\string##1}}}
\nolabels
%
\global\newcount\secno \global\secno=0
\global\newcount\meqno \global\meqno=1
\def\newsec#1{\global\advance\secno by1\message{(\the\secno. #1)}
\global\subsecno=0\eqnres@t\noindent{\bf\the\secno. #1}
\writetoca{{\secsym} {#1}}\par\nobreak\medskip\nobreak}
\def\eqnres@t{\xdef\secsym{\the\secno.}\global\meqno=1\bigbreak\bigskip}
\def\sequentialequations{\def\eqnres@t{\bigbreak}}\xdef\secsym{}
\global\newcount\subsecno \global\subsecno=0
\def\subsec#1{\global\advance\subsecno by1\message{(\secsym\the\subsecno. #1)}
\ifnum\lastpenalty>9000\else\bigbreak\fi
\noindent{\it\secsym\the\subsecno. #1}\writetoca{\string\quad 
{\secsym\the\subsecno.} {#1}}\par\nobreak\medskip\nobreak}
\def\appendix#1#2{\global\meqno=1\global\subsecno=0\xdef\secsym{\hbox{#1.}}
\bigbreak\bigskip\noindent{\bf Appendix #1. #2}\message{(#1. #2)}
\writetoca{Appendix {#1.} {#2}}\par\nobreak\medskip\nobreak}
%
%
\def\eqnn#1{\xdef #1{(\secsym\the\meqno)}\writedef{#1\leftbracket#1}%
\global\advance\meqno by1\wrlabeL#1}
\def\eqna#1{\xdef #1##1{\hbox{$(\secsym\the\meqno##1)$}}
\writedef{#1\numbersign1\leftbracket#1{\numbersign1}}%
\global\advance\meqno by1\wrlabeL{#1$\{\}$}}
\def\eqn#1#2{\xdef #1{(\secsym\the\meqno)}\writedef{#1\leftbracket#1}%
\global\advance\meqno by1$$#2\eqno#1\eqlabeL#1$$}
%
\newskip\footskip\footskip14pt plus 1pt minus 1pt 
\def\footnotefont{\ninepoint}\def\f@t#1{\footnotefont #1\@foot}
\def\f@@t{\baselineskip\footskip\bgroup\footnotefont\aftergroup\@foot\let\next}
\setbox\strutbox=\hbox{\vrule height9.5pt depth4.5pt width0pt}
\global\newcount\ftno \global\ftno=0
\def\foot{\global\advance\ftno by1\footnote{$^{\the\ftno}$}}
%
\newwrite\ftfile   
\def\footend{\def\foot{\global\advance\ftno by1\chardef\wfile=\ftfile
$^{\the\ftno}$\ifnum\ftno=1\immediate\openout\ftfile=foots.tmp\fi%
\immediate\write\ftfile{\noexpand\smallskip%
\noexpand\item{f\the\ftno:\ }\pctsign}\findarg}%
\def\footatend{\vfill\eject\immediate\closeout\ftfile{\parindent=20pt
\centerline{\bf Footnotes}\nobreak\bigskip\input foots.tmp }}}
\def\footatend{}
%
%
\global\newcount\refno \global\refno=1
\newwrite\rfile
\def\ref{[\the\refno]\nref}
\def\nref#1{\xdef#1{[\the\refno]}\writedef{#1\leftbracket#1}%
\ifnum\refno=1\immediate\openout\rfile=refs.tmp\fi
\global\advance\refno by1\chardef\wfile=\rfile\immediate
\write\rfile{\noexpand\item{#1\ }\reflabeL{#1\hskip.31in}\pctsign}\findarg}
\def\findarg#1#{\begingroup\obeylines\newlinechar=`\^^M\pass@rg}
{\obeylines\gdef\pass@rg#1{\writ@line\relax #1^^M\hbox{}^^M}%
\gdef\writ@line#1^^M{\expandafter\toks0\expandafter{\striprel@x #1}%
\edef\next{\the\toks0}\ifx\next\em@rk\let\next=\endgroup\else\ifx\next\empty%
\else\immediate\write\wfile{\the\toks0}\fi\let\next=\writ@line\fi\next\relax}}
\def\striprel@x#1{} \def\em@rk{\hbox{}} 
\def\lref{\begingroup\obeylines\lr@f}
\def\lr@f#1#2{\gdef#1{\ref#1{#2}}\endgroup\unskip}
\def\semi{;\hfil\break}
\def\addref#1{\immediate\write\rfile{\noexpand\item{}#1}} 
\def\startrefs#1{\immediate\openout\rfile=refs.tmp\refno=#1}
\def\xref{\expandafter\xr@f}\def\xr@f[#1]{#1}
\def\refs#1{\count255=1[\r@fs #1{\hbox{}}]}
\def\r@fs#1{\ifx\und@fined#1\message{reflabel \string#1 is undefined.}%
\nref#1{need to supply reference \string#1.}\fi%
\vphantom{\hphantom{#1}}\edef\next{#1}\ifx\next\em@rk\def\next{}%
\else\ifx\next#1\ifodd\count255\relax\xref#1\count255=0\fi%
\else#1\count255=1\fi\let\next=\r@fs\fi\next}
%

%
\newwrite\ffile\global\newcount\figno \global\figno=1
\def\fig{fig.~\the\figno\nfig}
\def\nfig#1{\xdef#1{fig.~\the\figno}%
\writedef{#1\leftbracket fig.\noexpand~\the\figno}%
\ifnum\figno=1\immediate\openout\ffile=figs.tmp\fi\chardef\wfile=\ffile%
\immediate\write\ffile{\noexpand\medskip\noexpand\item{Fig.\ \the\figno. }
\reflabeL{#1\hskip.55in}\pctsign}\global\advance\figno by1\findarg}
\def\xfig{\expandafter\xf@g}\def\xf@g fig.\penalty\@M\ {}
\def\figs#1{figs.~\f@gs #1{\hbox{}}}
\def\f@gs#1{\edef\next{#1}\ifx\next\em@rk\def\next{}\else
\ifx\next#1\xfig #1\else#1\fi\let\next=\f@gs\fi\next}
\newwrite\lfile
{\escapechar-1\xdef\pctsign{\string\%}\xdef\leftbracket{\string\{}
\xdef\rightbracket{\string\}}\xdef\numbersign{\string\#}}

\def\writestop{\def\writestoppt{\immediate\write\lfile{\string\pageno%
\the\pageno\string\startrefs\leftbracket\the\refno\rightbracket%
\string\def\string\secsym\leftbracket\secsym\rightbracket%
\string\secno\the\secno\string\meqno\the\meqno}\immediate\closeout\lfile}}
\def\writestoppt{}\def\writedef#1{}
\def\seclab#1{\xdef #1{\the\secno}\writedef{#1\leftbracket#1}\wrlabeL{#1=#1}}
\def\subseclab#1{\xdef #1{\secsym\the\subsecno}%
\writedef{#1\leftbracket#1}\wrlabeL{#1=#1}}
\newwrite\tfile \def\writetoca#1{}
\def\leaderfill{\leaders\hbox to 1em{\hss.\hss}\hfill}
\def\writetoc{\immediate\openout\tfile=toc.tmp 
   \def\writetoca##1{{\edef\next{\write\tfile{\noindent ##1 
   \string\leaderfill {\noexpand\number\pageno} \par}}\next}}}

%
\catcode`\@=12 
%
\edef\tfontsize{\ifx\answ\bigans scaled\magstep3\else scaled\magstep4\fi}
\font\titlerm=cmr10 \tfontsize \font\titlerms=cmr7 \tfontsize
\font\titlermss=cmr5 \tfontsize \font\titlei=cmmi10 \tfontsize
\font\titleis=cmmi7 \tfontsize \font\titleiss=cmmi5 \tfontsize
\font\titlesy=cmsy10 \tfontsize \font\titlesys=cmsy7 \tfontsize
\font\titlesyss=cmsy5 \tfontsize \font\titleit=cmti10 \tfontsize
\skewchar\titlei='177 \skewchar\titleis='177 \skewchar\titleiss='177
\skewchar\titlesy='60 \skewchar\titlesys='60 \skewchar\titlesyss='60
\def\titlefont{\def\rm{\fam0\titlerm}
\textfont0=\titlerm \scriptfont0=\titlerms \scriptscriptfont0=\titlermss
\textfont1=\titlei \scriptfont1=\titleis \scriptscriptfont1=\titleiss
\textfont2=\titlesy \scriptfont2=\titlesys \scriptscriptfont2=\titlesyss
\textfont\itfam=\titleit \def\it{\fam\itfam\titleit}\rm}
 \ifx\answ\bigans\else scaled\magstep1\fi
\ifx\answ\bigans\def\abstractfont{\tenpoint}\else
\font\abssl=cmsl10 scaled \magstep1
\font\absrm=cmr10 scaled\magstep1 \font\absrms=cmr7 scaled\magstep1
\font\absrmss=cmr5 scaled\magstep1 \font\absi=cmmi10 scaled\magstep1
\font\absis=cmmi7 scaled\magstep1 \font\absiss=cmmi5 scaled\magstep1
\font\abssy=cmsy10 scaled\magstep1 \font\abssys=cmsy7 scaled\magstep1
\font\abssyss=cmsy5 scaled\magstep1 \font\absbf=cmbx10 scaled\magstep1
\skewchar\absi='177 \skewchar\absis='177 \skewchar\absiss='177
\skewchar\abssy='60 \skewchar\abssys='60 \skewchar\abssyss='60
\def\abstractfont{\def\rm{\fam0\absrm}
\textfont0=\absrm \scriptfont0=\absrms \scriptscriptfont0=\absrmss
\textfont1=\absi \scriptfont1=\absis \scriptscriptfont1=\absiss
\textfont2=\abssy \scriptfont2=\abssys \scriptscriptfont2=\abssyss
\textfont\itfam=\bigit \def\it{\fam\itfam\bigit}\def\footnotefont{\tenpoint}%
\textfont\slfam=\abssl \def\sl{\fam\slfam\abssl}%
\textfont\bffam=\absbf \def\bf{\fam\bffam\absbf}\rm}\fi
\def\tenpoint{\def\rm{\fam0\tenrm}
\textfont0=\tenrm \scriptfont0=\sevenrm \scriptscriptfont0=\fiverm
\textfont1=\teni  \scriptfont1=\seveni  \scriptscriptfont1=\fivei
\textfont2=\tensy \scriptfont2=\sevensy \scriptscriptfont2=\fivesy
\textfont\itfam=\tenit \def\it{\fam\itfam\tenit}\def\footnotefont{\ninepoint}%
\textfont\bffam=\tenbf \def\bf{\fam\bffam\tenbf}\def\sl{\fam\slfam\tensl}\rm}
\font\ninerm=cmr9 \font\sixrm=cmr6 \font\ninei=cmmi9 \font\sixi=cmmi6 
\font\ninesy=cmsy9 \font\sixsy=cmsy6 \font\ninebf=cmbx9 
\font\nineit=cmti9 \font\ninesl=cmsl9 \skewchar\ninei='177
\skewchar\sixi='177 \skewchar\ninesy='60 \skewchar\sixsy='60 
\def\ninepoint{\def\rm{\fam0\ninerm}
\textfont0=\ninerm \scriptfont0=\sixrm \scriptscriptfont0=\fiverm
\textfont1=\ninei \scriptfont1=\sixi \scriptscriptfont1=\fivei
\textfont2=\ninesy \scriptfont2=\sixsy \scriptscriptfont2=\fivesy
\textfont\itfam=\ninei \def\it{\fam\itfam\nineit}\def\sl{\fam\slfam\ninesl}%
\textfont\bffam=\ninebf \def\bf{\fam\bffam\ninebf}\rm} 
%
%

\hyphenation{anom-aly anom-alies coun-ter-term coun-ter-terms}
\def\inv{^{\raise.15ex\hbox{${\scriptscriptstyle -}$}\kern-.05em 1}}

\def\Dsl{\,\raise.15ex\hbox{/}\mkern-13.5mu D} 
\def\dsl{\raise.15ex\hbox{/}\kern-.57em\partial}

\font\bigit=cmti10 scaled \magstep1
\def\lspace{\ifx\answ\bigans{}\else\qquad\fi}
\def\lbspace{\ifx\answ\bigans{}\else\hskip-.2in\fi} 
\def\boxeqn#1{\vcenter{\vbox{\hrule\hbox{\vrule\kern3pt\vbox{\kern3pt
        \hbox{${\displaystyle #1}$}\kern3pt}\kern3pt\vrule}\hrule}}}
\def\mbox#1#2{\vcenter{\hrule \hbox{\vrule height#2in
                \kern#1in \vrule} \hrule}}  
%
 \def\CO{{\cal O}} 

\def\vev#1{\langle #1 \rangle}

\def\darr#1{\raise1.5ex\hbox{$\leftrightarrow$}\mkern-16.5mu #1}

\def\roughly#1{\raise.3ex\hbox{$#1$\kern-.75em\lower1ex\hbox{$\sim$}}}

\def\OMIT#1{}

\def\eg{{\it e.g.}}
\def\etal{{\it et al.\/}}
\def\spur{\raise.15ex\hbox{/}\kern-.57em }

\def\CO{{\cal O}}
\def\ccdot{\hbox{\kern-.1em$\cdot$\kern-.1em}}
\def\frac#1#2{{#1\over#2}}

\def\larr#1{\raise1.5ex\hbox{$\leftarrow$}\mkern-16.5mu #1}

%
%
\def\fBs{{f^{\vphantom{\dagger}}_{B^*}}}
\def\fB{{f^{\vphantom{\dagger}}_{B}}}

\def\fD{{f^{\vphantom{\dagger}}_{D}}}
\def\gbb{{g^{\vphantom{\dagger}}_{B^*B\pi}}}
\def\gdd{{g^{\vphantom{\dagger}}_{D^*D\pi}}}
\def\mB{{m^{\vphantom{\dagger}}_B}}

\def\mD{{m^{\vphantom{\dagger}}_D}}
\def\mDstar{{m^{\vphantom{\dagger}}_{D^*}}}
%
%

\def\lsl{\spur {\kern0.1em l}}

%

%

\def\np#1#2#3{\NP{\bf #1} (#2) #3}
\def\pl#1#2#3{\PL{\bf #1} (#2) #3}
\def\prl#1#2#3{\PRL{\bf #1} (#2) #3}
\def\pr#1#2#3{\PR{\bf #1} (#2) #3}
\def\physrev#1#2#3{\PR{\bf #1} (#2) #3}

\def\blankref#1#2#3{   {\bf #1} (#2) #3}

\def\NP{{\it Nucl.\ Phys.\ }}
\def\PL{{\it Phys.\ Lett.\ }}
\def\PR{{\it Phys.\ Rev.\ }}

\def\PRL{{\it Phys.\ Rev.\ Lett.\ }}

%
%
%
\catcode`\@=11 
\global\newcount\exerno \global\exerno=0
\def\exercise#1{\begingroup\global\advance\exerno by1%
\medbreak
\baselineskip=10pt plus 1pt minus 1pt
\parskip=0pt
\hrule\smallskip\noindent{\sl Exercise \the\secno.\the\exerno \/}
{\ninepoint #1}\smallskip\hrule\medbreak\endgroup}
%
%
\def\newsec#1{\global\advance\secno by1\message{(\the\secno. #1)}
\global\exerno=0%
\global\subsecno=0\eqnres@t\noindent{\bf\the\secno. #1}
\writetoca{{\secsym} {#1}}\par\nobreak\medskip\nobreak}
\def\appendix#1#2{\global\exerno=0%
\global\meqno=1\global\subsecno=0\xdef\secsym{\hbox{#1.}}
\bigbreak\bigskip\noindent{\bf Appendix #1. #2}\message{(#1. #2)}
\writetoca{Appendix {#1.} {#2}}\par\nobreak\medskip\nobreak}
\catcode`\@=12 
%
%
\def\INSERTFIG#1#2#3{\vbox{\vbox{\hfil\epsfbox{#1}\hfill}%
{\narrower\noindent%
\multiply\baselineskip by 3%
\divide\baselineskip by 4%
{\ninerm Figure #2 }{\ninesl #3 \medskip}}
}}%
\def\yesanswer{y}
\def\PRLtype{n}

\def\ABSTRACTPACS#1#2{#1\vskip0.2in\centerline{#2}}
\Title{\vbox{\hbox{UCSD/PTH 94-27}\hbox{hep-ph/9412324}}}{\vbox{%
\centerline{Constraints on Form Factors For}
\centerline{Exclusive Semileptonic Heavy to Light Meson Decays}}}
\centerline{C. Glenn Boyd\footnote{$^{\ast}$}{gboyd@ucsd.edu},
Benjam\'\i n Grinstein\footnote{$^{\dagger}$}{bgrinstein@ucsd.edu} and
Richard F. Lebed\footnote{$^{\ddagger}$}{rlebed@ucsd.edu}}
\bigskip\centerline{Department of Physics}
\centerline{University Of California, San Diego}
\centerline{La Jolla, California 92093-0319}
%
%
\vskip .3in
\def\ffsdefd{(1.1)}
\def\bndfp{(1.2)}
\def\meiman{[1]}
\def\bourrely{[2]}
\def\rtaron{[3]}
\def\bunch{[4]}
\def\grinmende{[5]}
\def\twopntfnctn{(2.1)}
\def\dsptnrltn{(2.2)}
\def\onegf{(2.3)}
\def\kdefd{(2.4)}
\def\ttoz{(2.5)}
\def\onegfb{(2.6)}
\def\phidefd{(2.7)}
\def\BOXX{fig.~1}
\def\ipdefd{(2.8)}
\def\fis{(2.9)}
\def\Idefd{(2.10)}
\def\foneftwo{(2.11)}
\def\vind{[6]}
\def\accmor{[7]}
\def\cleobr{[8]}
\def\brbunch{[9]}
\def\hqcpt{[10]}
\def\letticefB{[11]}
\def\HEMCGC{[12]}
\def\WSB{[13]}
\def\IWpole{[14]}
\def\ISGWtwo{[15]}
\def\DboundA{fig.~2}
\def\expfd{[16]}
\def\etal{{\it et al.}}
\def\G{{\Gamma}}
\def\t{{\theta}}
\def\b{{\beta}}%

\def\chiTL{{\chi^{\vphantom{\dagger}}_{T,L}}}
\def\semi{{;}}

\ABSTRACTPACS{
We use rigorous QCD dispersion relations to derive model-independent
bounds on the $\overline B \to \pi l \overline \nu$, $D \to \pi
\overline l \nu$ and $D \to \overline K \, \overline l \nu$ form
factors.  These bounds are particularly restrictive when the value of
the observable form factor at one or more kinematic points is assumed.
With reasonable assumptions
 we find
$\fB \leq 195 \, \rm{MeV}$
and  that the shape of
the form factor becomes severely constrained. These constraints are
useful both for model discrimination and for model-insensitive
extraction of CKM mixing parameters.
}{PACS: 13.20He, 13.20Fc, 12.39.-x, 12.40.Vv}
%
%
\ifx\PRLtype\yesanswer
\else
\Date{December 1994}
\fi
\newsec{ Introduction}
Charmless $B$-meson decays are of great interest because the rate
depends directly on a fundamental parameter, the CKM matrix element
$|V_{ub}|$. Its determination requires knowledge of non-perturbative
hadronic matrix elements. Semileptonic decays involve the hadronic
matrix element of a partially conserved current, and there is hope
that one may calculate them, or at least model them, with some
precision. The inclusive charmless decay rate is only
measured at $E_e\approx E_{e,{\rm max}}$, where the theoretical calculation is
highly uncertain.  Alternatively one can measure
exclusive rates, \eg, $d\G(\bar B\to\pi e\bar \nu)$, over the whole kinematic
range. One then needs theoretical calculations of the hadronic matrix elements.
Similarly, $D\to \bar K $ ($D\to \pi$) semileptonic decays are
interesting because they allow determination of $|V_{cs}|$
($|V_{cd}|$).

In this letter we show that one can calculate rather good bounds on
the rates for semileptonic exclusive $B$ and $D$ decays to light
pseudoscalar mesons. Parametrizing the $B \pi$ matrix element of the
flavor-changing vector current $V_\mu=\bar u\gamma_\mu b$ by
\eqn\ffsdefd{
\vev{\pi(p')| V_\mu |\bar B(p)} = f_+(q^2)(p+p')_\mu +
        f_-(q^2)(p-p')_\mu ~,
}
we obtain inequalities of form
\eqn\bndfp{
F_-(q^2)\le |f_+(q^2)| \le F_+(q^2)~.}
Below we describe the calculation of the functions $F_\pm$.  The bounds are
model independent. They involve a few physical parameters:
masses, decay constants and the $B^*$-$B$-$\pi$
coupling $\gbb$, that must be determined independently. In addition, for a
strong bound one needs the value of the form factor for at least one kinematic
point, but this may not  require
additional parameters.

The method we will employ is not new\ref\meiman{
        N. N. Meiman, Sov. Phys. JETP \blankref{17}{1963}{830}\semi
        S. Okubo and I. Fushih, \pr{D4}{1971}{2020}\semi
        V. Singh and A.K. Raina, {\it Fortschritte der Physik}
        \blankref{27}{1979}{561}}.
It was used to obtain bounds on
form factors for semileptonic $K$-meson decays\ref\bourrely{
      C. Bourrely, B. Machet and E. de Rafael, \np{B189}{1981}{157}}.
The method has also
been applied to the decay $B\to \bar D e \nu$%
\ref\rtaron{E. de~Rafael and J. Taron, \pl{B282}{1992}{215};
\physrev{D50}{1994}{373}}, but here there is an
important difference\nref\bunch{E. Carlson,
J. Milana, N. Isgur, T. Mannel, and W. Roberts,
       \pl{299}{1993}{133}\semi
       A. Falk, M. Luke, and M. Wise, \pl{299}{1993}{123}\semi
       J. K\"orner and D. Pirjol, \pl{301}{1993}{257}}%
\nref\grinmende{B. Grinstein and P. Mende, \pl{299}{1993}{127}}%
\refs{\bunch,\grinmende}. While there are no
poles below the onset of ${\it vacuum}\to
\bar K \pi$,  there are several resonances with masses smaller than $\mB+\mD$,
namely, the onium-like $B_c$'s.  As pointed out in Ref.~\grinmende,
the case $\bar B\to\pi e \bar \nu$ is intermediate between these:
there is exactly one resonance below the onset of the $\bar B$-$\pi$
continuum, the $\bar B^*$. This is phenomenologically true. It is also
guaranteed in the heavy quark limit for $m_\pi$ small and fixed, since
the $B^*$-$B$ mass splitting is $\CO(1/\mB)$. In the case $B\to \bar
D$ the multitude of resonances below $\mB+\mD$ renders the method
quite weak, even though heavy quark symmetries fix the values of the
form factors at one kinematic point. For $B\to\pi$ the situation is
improved because there is only one such resonance.

\newsec{Method}
The derivation of the bounds is well known. We present a short version
here both to establish notation and to underline where we may deviate
from the standard case.
Consider the two-point function
\ifx\PRLtype\yesanswer
\eqnn\twopntfnctn
$$\eqalignno{
i \int d^4\!x e^{iqx}\vev{{\rm T} V_\mu(x) V^\dagger_\nu(0)}
=(q_\mu &q_\nu-q^2 g_{\mu\nu})\Pi_T(q^2)&\cr
& + g_{\mu\nu}\Pi_L(q^2).&\twopntfnctn\cr
}$$
\else
\eqn\twopntfnctn{
i \int d^4\!x e^{iqx}\vev{{\rm T} V_\mu(x) V^\dagger_\nu(0)}
=(q_\mu q_\nu-q^2g_{\mu\nu})\Pi_T(q^2)
 + g_{\mu\nu}\Pi_L(q^2).
}
\fi
In QCD the structure functions satisfy
a once-subtracted dispersion relation:
\eqn\dsptnrltn{
\chiTL (Q^2)=\left.{{\partial\Pi_{T,L}}
\over{\partial q^2}}\right|_{q^2=-Q^2}=
{1\over\pi}\int_0^\infty dt \, {{{\rm Im}\,\Pi_{T,L}(t)}\over{(t+Q^2)^2}}
.}
The absorptive parts ${\rm Im}\,\Pi_{T,L}(q^2)$ are obtained by
inserting real states between the two currents on the right-hand side
of Eq.~\twopntfnctn. A judicious choice of $\mu$ and $\nu$ makes this
a sum of positive definite terms, so one can obtain strict
inequalities by concentrating on the term with intermediate states
of $\bar B\pi$
pairs.  For $Q^2$ far from the resonance region the two-point
function can be computed reliably from perturbative QCD. In
particular, for large $b$ quark mass, $Q^2=0$ is far from resonances.
One resulting inequality of this method is
\eqn\onegf{
\int_{t_+}^\infty dt\, k(t) |f_+(t)|^2 \le 1,
}
where, neglecting the light quark mass,
\def\onethird{{\textstyle{1\over3}}}
\eqn\kdefd{
k(t)= {\onethird}(m_b/t)^{2}[(1-t_+/t)(1-t_-/t)]^{3/2},
}
and $t_\pm=(\mB\pm m_\pi)^2$.

Using knowledge of the analytic structure of the form factor plus
the bound Eq.~\onegf\ one can derive bounds\refs{\meiman,\bourrely}\
on the form factor in the physical region of semileptonic decay,
$0\le t \le t_-$. To this end we map the complex $t$-plane onto the
unit disk $|z|\le1$ by the transformation
\eqn\ttoz{
\sqrt{{t_+-t\over t_+-t_-}}={{1+z}\over{1-z}}~.
}
The two branches of the root for $t_+\le t$ are mapped into the unit
circle $z=e^{i\theta}$, while the regions $ t \le t_-$ and $t_-\le t<
t_+$ are mapped into the segments of the real axis $-1< z\le0$ and
$0\le z<1$, respectively. In terms of this new variable the inequality
\onegf\ is
\eqn\onegfb{
{1\over2}\int_0^{2\pi} d\t\, w(\t) |f_+|^2 \le1,
}
where
$w(\t) = k(t(\t)) {{dt}\over{d\t}}$.
Next we construct a function $\phi(z)$ analytic in $|z|<1$ such that
$|\phi(e^{i\theta})|^2=w(\theta)$:
\eqn\phidefd{
\phi(z)= {2^{5/2}m_b\over\sqrt3} {  (t_+-t_-)^{-1/2}(1+z)^{2}\over
(1-z)^{9/2}} \left(\beta_+ + {{1+z}\over{1-z}} \right)^{\!\!-5}\!\!,
}
where $\b_+= \sqrt{t_+/(t_+-t_-)}$.

\gdef\CAPone{Upper and lower bounds (solid lines) on $f_+(t)$
for $B\to\pi$, plotted against ${t/ M_B^2}$.  The ``pure pole''
form factor $f_{\rm pole}(t)$ is plotted in dashed lines, while the
WSB model is in dot-dashed lines.  The bound is from the (a)
$2\times2$, (b) $3\times3$ and (c-d) $4\times4$ determinants.  In (b)
we assume $f_+(t_-)\simeq f_{pole}(t_-)$. In (c) we use $f_+(0)$ from
the WSB model and $f_+(t_-)$ from $B^*$ pole dominance. In (d) we use
as inputs $f_+(t_-)$ and $f_+(t_--2\mB m_\pi)$ from the pole dominance
assumption of heavy meson chiral perturbation theory. At the scale of
the figure the bounds are indistinguishable.}
\nfig\BOXX{\CAPone}

\ifx\PRLtype\yesanswer
\else
\vskip0.7truein
\def\epsfsize#1#2{0.80#1}%
\INSERTFIG{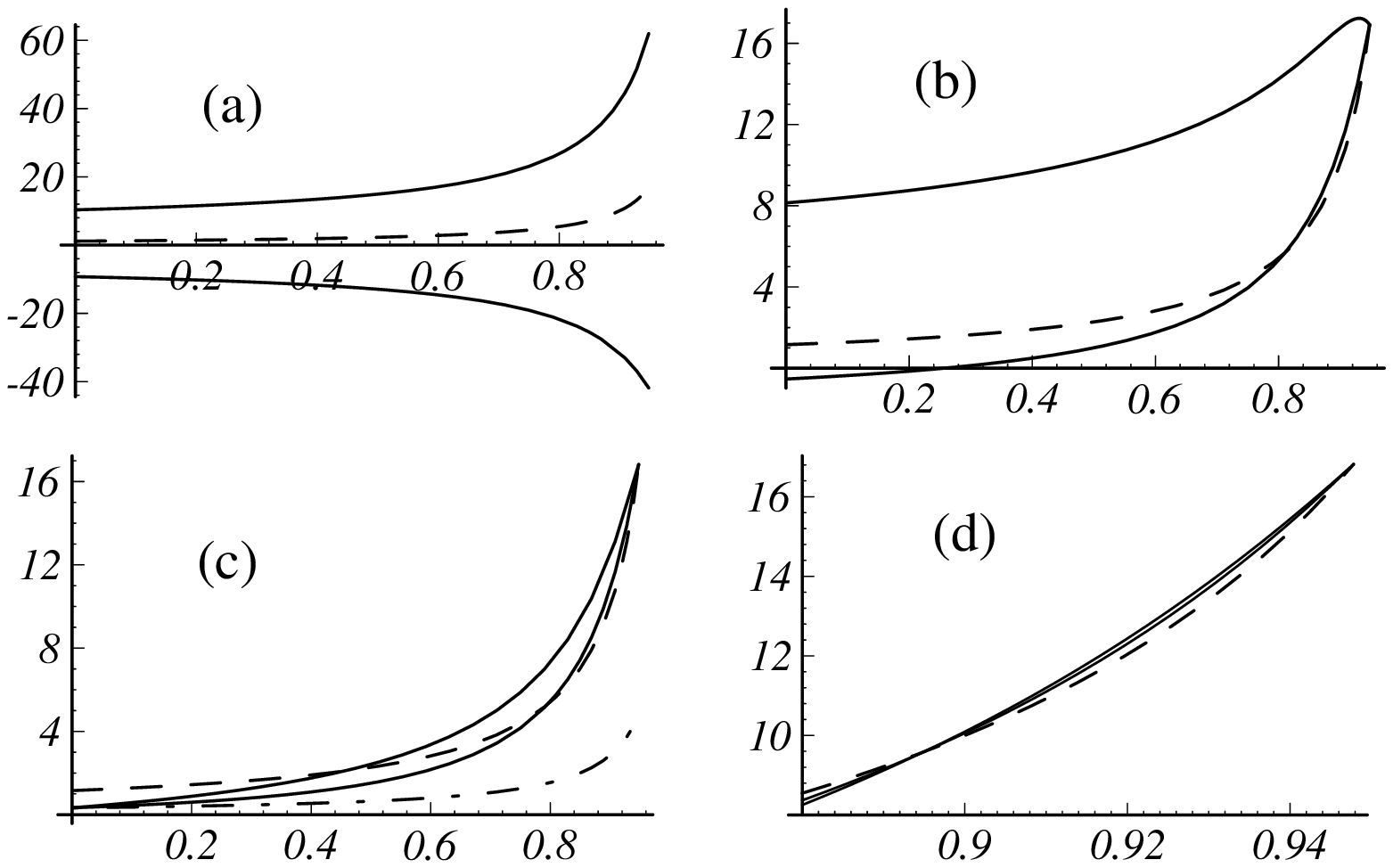}{1. }{\CAPone}
\fi

With Ref.~\bourrely, let us define an inner product on the space of
complex functions of a real variable $\theta$, with $0\le\theta<2\pi$,
by
\eqn\ipdefd{
(f,g)\equiv{1\over2\pi}\int_0^{2\pi}d\theta\, f^*(\theta)g(\theta).
}
Next let
\eqn\fis{
\eqalign{
f_0(\t) &= \phi(e^{i\t}) f_+(e^{i\t})~,\cr
f_i(\t) &= {1\over{1-\bar z_i e^{i\t}}}~,\cr}
}
where $z_i$ are arbitrary complex numbers with $|z_i|<1$.
With this, we have
\eqn\Idefd{
I\equiv (f_0,f_0)={1\over2\pi}\int_0^{2\pi} d\t\, w(\t)
|f_+|^2\le{1\over\pi}~.
}
Using Cauchy's theorem we can evaluate the other inner products.  We
must bear in mind that the form factor $f_+$ has a pole at $t= t_*
\equiv m_{B^*}^2$, corresponding to $z_*=z(t=m_{B^*}^2)$ inside the
unit circle. For example,
\eqn\foneftwo{
(f_1,f_0) =  \phi(z_1)f_+(z_1) + {\left.\hbox{Res}\,(\phi
f_+)\right|_{z_*}\over z_*-z_1 }.}
\ifx\PRLtype\yesanswer
\enddoublecolumns
\vskip0.9truein
\def\epsfsize#1#2{0.80#1}%
\INSERTFIG{fig1.eps}{1. }{\CAPone}
\vfill\eject
\begindoublecolumns
\else
\fi
\noindent From the positivity of the inner product we have that the
matrix $(f_i,f_j)$ has positive determinant. \OMIT{This gives
\eqn\posdet{
{\rm det}\pmatrix{I&g(z_1)\cr
                  g^*(z_1)&{1\over 1-|z_1|^2}\cr}
          \ge0
\quad,\quad
{\rm det}\pmatrix{I&g(z_1)&g(z_2)\cr
                  g^*(z_1)&{1\over 1-|z_1|^2}&{1\over 1-\bar z_1 z_2}\cr
                  g^*(z_2) & {1\over 1-\bar z_2 z_1} &{1\over 1-|z_2|^2}\cr}
          \ge0,
          }
and so on. }
Inequalities \bndfp\ follow; it is straightforward to display
analytic expressions for the bounding functions $F_\pm$.
We can further improve our bounds by including the vector meson
contribution to the absorptive part of the structure functions, and by
generalizing the calculation to nonzero $Q^2$.

As a side benefit we find that
$\fBs < {1\over 4 \pi}\sqrt{3 \over 2}
           { m_{B^*}^3 \over m_b}$.
This is consistent with the heavy quark symmetry relation
$\fBs = (\mB\mD)^{1/2} \fD $ and the bound
$\fD < {\mD\over 4 \pi}$ from an analogous dispersion
relation\ref\vind{V.A. Novikov \etal, \prl{38}{1977}{626}\semi
    S. Narison and E. de Rafael, \np{169}{1980}{253}\semi
    S. Narison, E. de Rafael, and F. Yndurain, Marseille
    preprint 80/P1186 (1980) (unpublished)}.

\newsec{Analysis and Discussion}
\subsec{$B\to\pi$}
The bounds on $f_+$ require explicit knowledge of the residue
$F_*=\fBs\gbb$. Heavy quark symmetries imply $\gbb/m_B=\gdd/m_D$ and
$\fBs=\mB\fB$, at leading
order. An experimental upper bound on the $D^*$ width%
\ref\accmor{The ACCMOR Collaboration (S.~Barlag \etal),
       \pl{278}{1992}{480}}, together with
measurements of the $D^*$ decay fractions%
\ref\cleobr{The Cleo Collaboration ((F. Butler,{\it et al.}),
     \prl{69}{1992}{2041}},
gives (using $90\%$ confidence values)\ref\brbunch{J. Amundson, \etal,
     \pl{296}{1992}{415}\semi
     H-Y Cheng \etal, \physrev{D47}{1993}{1030}\semi
     P. Cho and H. Georgi, \pl{B296}{1992}{408}; erratum {\it ibid.} {\bf
     B300}(1993)410} $0.06\le g^2\le 0.5$,
where $g=f_\pi\gdd/\mD$ to leading order in heavy meson chiral
perturbation theory\ref\hqcpt{
    M.~Wise, \physrev{D45}{1992}{2188}\semi
    G.~Burdman and J.~F.~Donoghue, \pl{280}{1992}{287}\semi
    T.~M.~Yan \etal, \physrev{D46}{1992}{1148}}.
Monte Carlo
simulations of quenched lattice QCD give
$\fB({\rm MeV})$ in the range\ref\letticefB{As.  Abada {\it et
al.}, \np{B376}{1992}{172}\semi
C.R.  Allton {\it et al.}, APE Collaboration, \pl{B326}{1994}{295}\semi
H. Wittig {\it et al.}, UKQCD Collaboration, {\it Nucl. Phys.\/}{\bf
B}(Proc. Suppl.) {\bf 34} (1993) 462\semi
C.W. Bernard {\it et al.}, \pr{D49}{1994}{2536}\semi
C.  Bernard, {\it et al.}, (MILC collaboration), WASH-U-HEP-94-37,
hep-lat/9411080}\ 150--290 with about 20\% errors, and an unquenched
calculation
gives\ref\HEMCGC{K. Bitar {\it et al.}, HEMCGC Collaboration,
\pr{D48}{1993}{370}}\ $200\pm48$. Clearly $F_*$ is poorly known.  In what
follows we shall take $F_*=33 \,\rm{GeV}^2$, corresponding to
$g^2=0.5$ and $\fB=220$~MeV.  Our bounds are stronger for larger
$F_*$, so the value we have chosen is not conservative, but rather
intended to illustrate the potential of the method.  We also take $Q^2
= -16~\rm{GeV}^2\!$, which is chosen to be closer to the resonance
region without violating our perturbative QCD assumption. This
typically narrows the band between the upper and lower bounds by
10--15\%.  The results of Ref.~\bourrely\ may be used to gauge the
reliability of this choice of $Q^2\!$. In addition, we include the
contribution of the $B^*$ to our dispersion relation, but the
resulting improvement is typically only a few percent.

Figure \xfig\BOXX{a} shows in solid lines the upper and lower
bounds from the $2\times2$ determinant. The abscissa in all $B$ meson plots is
presented in units of ${t/ m_B^2}$.
For reference we have plotted in dashed
lines a ``pure pole'' form factor $f_{pole}(t) = F_*/(m_{B^*}^2-t)$.
Although not very stringent, this bound uses the minimal set of
assumptions and could be used to put a rigorous lower bound on
$|V_{ub}|$ from a measurement of the width of $\bar B\to\pi e \bar
\nu$.

Bounds using the value of $f_+$ at one or more points are
significantly more restrictive. The proximity of the $B^*$ pole to the
region of maximum momentum transfer suggests $f_+(t_-)=f_{pole}(t_-)$
to good approximation.  We make this assumption in Fig.~\xfig\BOXX{b},
which requires a $3\times3$ determinant.  The dashed line shows the
simple pole curve, which, remarkably, falls outside the region allowed
by our bounds for values of momentum transfer close to $t_-$.  We find
one must decrease the value of $F_*$ to 23 GeV$^2$ before the pole
term lies entirely within the allowed region.

There are several models for $f_+$ in the literature. They are
intended to give a numerical approximation to the actual form factor
in the physical region for $\bar B \to\pi e\bar \nu$. One can test
whether a particular model is consistent with QCD by using an
arbitrary number of points $f_+(t_i)$ in our bounds. We will content
ourselves with bounds that use the value of $f_+$ at two points. This
requires a computation with a $4\times4$
determinant. We take $f_+(t_-) =
f_{pole}(t_-)$ as above, and fix a second point $f_+(t_i)$ from the
model under scrutiny.

The model of Wirbel, Stech and Bauer (WSB) has $f_+(0)=0.33$, and
assumes a single $B^*$-pole shape\ref\WSB{M. Wirbel, B. Stech and M.
Bauer, {\it Z. Phys.\/}
\blankref{C29}{1985}{637}}.  Presumably it is not intended to describe the
form factor accurately as $t\to t_-$\ref\IWpole{N. Isgur and M.B.
Wise, \physrev{D41}{1990}{151}}. Figure~\xfig\BOXX{c}\
shows the bounds obtained using $f_+(0)$ from this model in solid
lines, the pure pole $ f_{pole}(t)$ in dashes, and the WSB model
prediction in dot-dashes.  For the given value of $F_*$,
WSB falls outside of our bounds over the entire physical range.  For
$F_*< 23 \, \rm{GeV}^2$, the WSB curve lies within the bounds over a
range from $t=0$ to some $t_{crit}$, where $t_{crit}$ increases as
$F_*$ decreases.  A revised version of the model of Isgur {\it et
al.}\ref\ISGWtwo{D. Scora and N.  Isgur, CEBAF-TH-94-14}\ gives a
somewhat smaller form
factor for $B\to \pi$, leading to a smaller value of $t_{crit}$.

\ifx\PRLtype\yesanswer
\def\epsfsize#1#2{0.50#1}%
\else
\def\epsfsize#1#2{0.60#1}%
\fi
\def\CAPtwo{Upper and lower bounds on $f_+(t)$ for
$D\to\pi$, assuming $f_+(t_-)\simeq F_*/(m_{D^*}^2-t_-)$. The curve
$F_*/(m_{D^*}^2-t)$ is shown as a dashed line. The abscissa is given
in units of ${t/ M_D^2}$.}
\nfig\DboundA{\CAPtwo}
\INSERTFIG{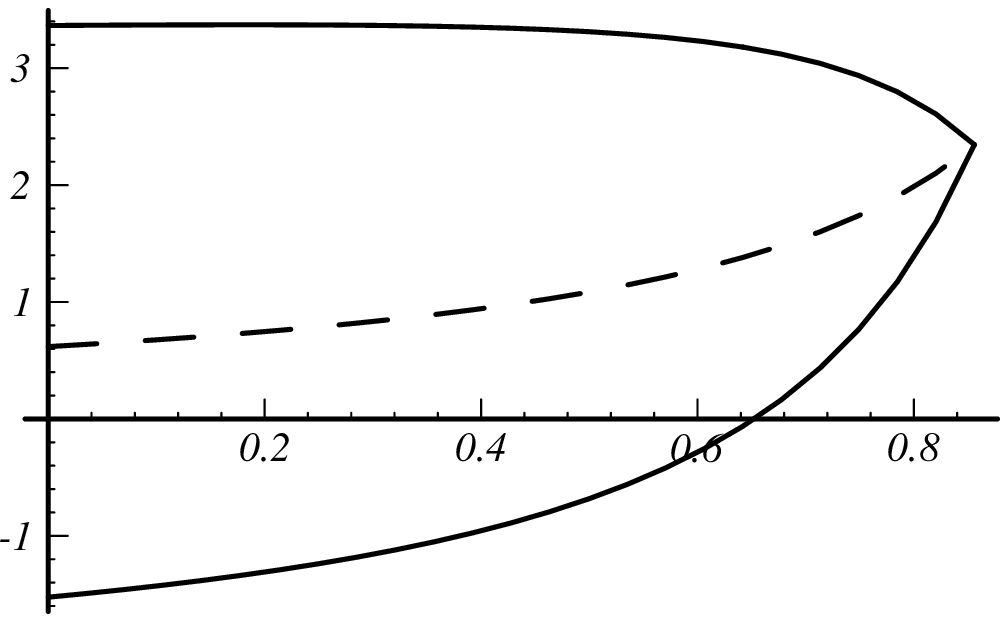}{2. }{\CAPtwo}

The validity of chiral perturbation theory for heavy mesons hinges on
single pole dominance of $f_+$ at and near $t=t_-$.
Figure~\xfig\BOXX{d} shows the bounds using input normalizations $f_+(t_-) =
f_{pole}(t_-)$ and $f_+(t_--2\mB m_\pi) = f_{pole}(t_--2\mB m_\pi)$.
This simply assumes heavy meson chiral perturbation theory is valid at
both $E_\pi=m_\pi$ and $2m_\pi$. The pure pole is again shown in
dashes. Either
the effects of higher resonances are non-negligible, or the value of
$F_*$ is inconsistent with chiral perturbation theory.  Insisting on
the validity of heavy meson chiral perturbation theory in this range
implies an upper bound, $F_*\le 10$ $ \rm{GeV}^2$.  Substituting the
lower bound in Ref.~\brbunch\ for $g^2$ then gives $\fB
\leq 195\, \rm{MeV}$.

\subsec{$D\to K$, $D\to\pi$}
For $D^+\to\pi^0$, $D^0\to\pi^-$ and
$D_s^+\to\pi^0$ we have  $\mDstar>\mD+m_\pi$ so we need no
{\it a priori} knowledge of the residue $F_*$ of the vector meson
pole.  However, useful bounds are obtained only if one has additional
information about the form factors.

Assuming the value of the form factor for $D^0\to\pi^-$ is dominated
by the $D^{*+}$ pole at $t=t_-\equiv (\mD-m_\pi)^2$ gives the bound in
Fig.~\xfig\DboundA. We have taken $Q^2 =0$ and $F_* = 2.5$ ${\rm
GeV}^2\!$, and plotted the pure pole in dashes.  A more restrictive
bound follows from using two normalization points, as in the $B$ meson
analysis. However, the perturbative QCD calculation is less reliable
than in the $B$ meson case.

The experimental measurements of $f^{{\vphantom{2}}}_{D_s}$\ref\expfd{
WA75 Collaboration, (S.Aoki \etal), {\it Prog. Theor. Phys.\/} {\bf
89} (1993) 131\semi The CLEO2 Collaboration, to be published\semi The
ARGUS Collaboration, (H. Albrecht \etal), {\it Z. Phys.\/} {\bf C54}
(1992) 1}\ are one to two standard deviations from the bound of
Ref.~\vind. How this bound eventually fares will shed light on the
minimal value of $Q^2$ consistent with reliable limits on $f_+^{D \to
\pi}(t)$.

\newsec{ Summary}
The analytic structure of form factors for heavy to light semileptonic
meson decays makes them well suited to analysis by simple dispersion
relations.  The validity of this analysis depends on the use of
perturbative QCD calculations at a distance $(M_{B^*,D^*}^2 + Q^2)$
from the resonance region.

For $\overline B \to \pi l \overline \nu$ decays, only the value of
the product of the decay constant $\fBs$ and the coupling
$\gbb$, $F_*=\fBs\gbb$, is necessary for model-independent bounds on
the experimentally accessible pion form factor $f_+(q^2)$. For $D \to
\pi \bar l \nu$ and $D \to \overline K \, \bar l \nu$ decays, even this
input is unnecessary. Together with experimental data, these form
factor bounds yield model-independent lower bounds on the
Cabbibo-Kobayashi-Maskawa parameters $|V_{ub}|$, $|V_{cs}|$ and
$|V_{cd}|$.

Much more restrictive form factor bounds result if the value of
$f_+(q^2)$ is known at a single kinematic point. This normalization
may come from experiment, lattice calculations, or phenomenological
and QCD-inspired
models. These form factor bounds allow the experimental extraction of
both upper and lower bounds on CKM angles, and place significant
restrictions on models. For example, using a one-point normalization,
we show that heavy meson chiral
perturbation theory with minimal particle content is inconsistent
for values of $F_* > 23$~GeV$^2$.

Using the normalization of $f_+(q^2)$ at two kinematic points yields
even more restrictive form factor bounds. Typically the shape of the
form factor between the normalization points is very severely
constrained.  This can be used to interpolate between models in
disparate regions of phase space, or to restrict the parameter space
of a given model. In the case of heavy meson chiral perturbation
theory, we find consistency only if $F_* < 10$~GeV$^2$. This
translates into the prediction $\fB < 195$~MeV.  Similar analyses
may be applied to other models.  We hope to present the
consequences of our bounds more thoroughly in a future work.

\vskip1.2cm
{\it Acknowledgements}\hfil\break
The research of one of us (B.G.)  is funded in part
by the Alfred P. Sloan Foundation. This work is supported in part by the
Department of Energy under contract DOE--FG03--90ER40546.

\footatend\vfill\supereject\immediate\closeout\rfile\writestoppt
\baselineskip=14pt\centerline{{\bf References}}\bigskip{\frenchspacing%
\parindent=20pt\escapechar=` \input refs.tmp\vfill\eject}\nonfrenchspacing
\bye